# Photoelectron track length distributions measured in a negative ion time projection chamber

Z. R. Prieskorn, J. E. Hill, P. E. Kaaret, and J. K. Black

*Abstract*– We report photoelectron track length distributions between 3 and 8 keV in gas mixtures of Ne+$CO_2$+$CH_3NO_2$ (260:80:10 Torr) and $CO_2$+$CH_3NO_2$ (197.5: 15 Torr). The measurements were made using a negative ion time projection chamber (NITPC) at the National Synchrotron Light Source (NSLS) at the Brookhaven National Laboratory (BNL). We report the first quantitative analysis of photoelectron track length distributions in a gas. The distribution of track lengths at a given energy is best fit by a lognormal distribution. A powerlaw distribution of the form, $f(E)=a(E/E_o)^n$, is found to fit the relationship between mean track length and energy. We find $n=1.29\pm0.07$ for Ne+$CO_2$+$CH_3NO_2$ and $n=1.20\pm0.09$ for $CO_2$+$CH_3NO_2$. Understanding the distribution of photoelectron track lengths in proportional counter gases is important for optimizing the pixel size and the dimensions of the active region in electron-drift time projection chambers (TPCs) and NITPC X-ray polarimeters.

*Index Terms*-photoelectron track length, negative ion time projection chamber, X-ray detectors, nitromethane, $CH_3NO_2$.

## I. INTRODUCTION

Imaging photoelectron tracks is the key to broadband, polarization sensitive X-ray detectors. When an X-ray photoionizes an atomic s-orbital, the photoelectron is ejected according to a $\sin^2\theta\cos^2\varphi$ probability distribution, where $\theta$ is the polar angle and $\varphi$ the polarization-dependent azimuthal angle [1]. If the photoelectron track is imaged with sufficient resolution the initial direction of the photoelectron can be determined and from imaging many tracks the polarization of the incident X-rays can be measured. Multiple groups have utilized this property to build X-ray polarimeters [2]–[7]. Photoelectron track length measurements can provide a diagnostic for optimizing the pixel size, gas pressure, and the dimension of the active region in an X-ray polarimeter. The two most important detector properties that ensure the track can be accurately measured are the pixel size and the dimension of the detector active region.

Photoelectron continuous slowing down approximation (CSDA) range has been measured extensively in foils [8]–[10] but measurements in gas mixtures have been more difficult to obtain due to the poor spatial resolution of proportional counters relative to the length of the tracks. Electron ranges have been estimated for a few select gases [11], [12] using multi-wire proportional counters. These experiments made estimates of the photoelectron range with a resolution of approximately 200 $\mu$m by taking an average of the delay between signals received on each end of the anode array in a position sensitive multi-wire proportional counter. Previous authors report the full-width at half maximum (FWHM) of the resolution measurement as a function of energy, which should scale with the range.

The dependence of the photoelectron range on energy was found to be well fitted by a power law with an index of ~1.8 in carbon gases at 750 Torr in the energy range of 5-25 keV [12]. In foils, a power law also provides the best fit with an index of ~1.3 [9]. For gas mixtures with a noble gas component the data do not fit the expected power-law distribution as the X-ray energy approaches the ionization energy of the atom. This is due to the high probability of an Auger electron occurring and at X-ray energies near the atoms ionization energy it will have kinetic energy comparable to that of the photoelectron. However, at higher X-ray energies the power-law behavior is recovered as the photoelectron energy begins to dominate. In previous measurements the photoelectron range or track length was only an average for each photoelectron energy. By imaging the photoelectron tracks with a NITPC, we have sufficient resolution to measure the length of each individual track. The definition of track length used in this experiment is the distance from the barycenter of charge at the interaction point to the barycenter of charge at the stop point and is described in more detail in §III.A. We have measured distributions of photoelectron track lengths, in two gas mixtures, as a function of X-ray energy and tested multiple models as a fit to the data. This is the first quantitative treatment of track length reported for a gas, although recently other authors have shown track length distributions without providing a quantitative analysis of the data [13]–[15].

In §II we describe the design of the NITPC detector and the experimental setup at the Brookhaven National Laboratory. §III is a description of how the measurements are made and how the track length is determined. §IV presents a discussion of our experimental results.

Manuscript submitted to arXiv June 17, 2014. This work was supported in part by NASA grants NNX08AF46G and NNX07AF21G.

Z. R. Prieskorn is with the Pennsylvania State University, University Park, PA 16802 USA (telephone: 319-400-1809, e-mail: prieskorn@psu.edu).
J. E. Hill is with the NASA Goddard Space Flight Center, Greenbelt MD 20771 USA (telephone: 301-286-0572, e-mail: joanne.e.hill@nasa.gov).
P. E. Kaaret is with the University of Iowa, Iowa City, IA 52240 USA (telephone: 319-335-1985, e-mail: Philip-kaaret@uiowa.edu).
J. K. Black is with Rock Creek Scientific, 1400 East-West Hwy, Suite 807, Silver Spring MD 20910 USA (e-mail: kevin.black@nasa.gov)

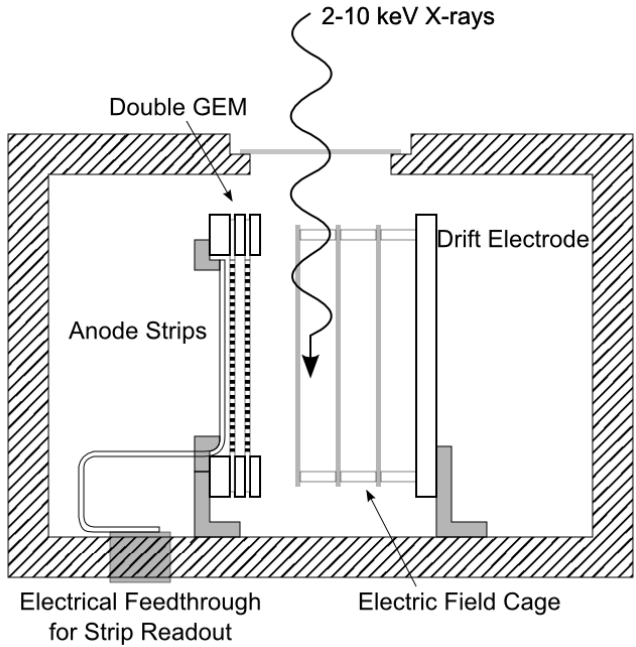

Figure 1. Diagram of the Negative Ion Time Projection (NITPC) Chamber used to make measurements of photoelectron tracks from low energy X-rays. The NITPC implementation is described further in §II.A.2).

## II. EXPERIMENTAL SETUP

### A. NITPC X-ray polarimeter

#### 1) Theory of operation

The theory of operation for the NITPC X-ray polarimeter is described in detail by other authors [4], [6], [7], [16]. The detector is shown schematically in Figure 1 and an isometric view is shown in Figure 2. The X-rays enter the polarimeter through a thin beryllium window. The active volume is defined by the drift electrode and the double gas electron multiplier (GEM). The drift electrode is set to a negative voltage relative to the voltage applied to the closest side of the GEM creating an electric field that will drift negative charges to the GEMs. A field-shaping cage maintains a uniform field (better than 1 %) between the drift electrode and GEM setup. Strip anodes are located directly behind the GEMs and the strips are aligned to the holes in the GEM. The strip pitch (0.121 mm) was chosen to make this possible. The GEM hole pitch is 0.140 mm and the hole layout is hexagonal, making the distance between rows of holes 0.121 mm.

An X-ray ionizes a gas atom in the active region of the detector. A photoelectron is ejected with kinetic energy equal to the X-ray energy minus the energy required to free the electron from the atom and in a direction related to the X-ray angle of polarization. The photoelectron loses energy via collisions with the surrounding gas particles as it travels through the gas. Some collisions result in gas atoms becoming ionized, creating electron/ion pairs. Nitromethane ($CH_3NO_2$) is used as a gas additive because it has an electron affinity and quickly captures the free electrons, forming negative ions [16]. The negative ions have a drift velocity two orders of magnitude slower than electrons. This provides two advantages: charge cloud diffusion is reduced and the power requirements for the electronics to read out the signal is decreased. The negative ions then drift toward the small holes of the gas electron multiplier (GEM) [17].

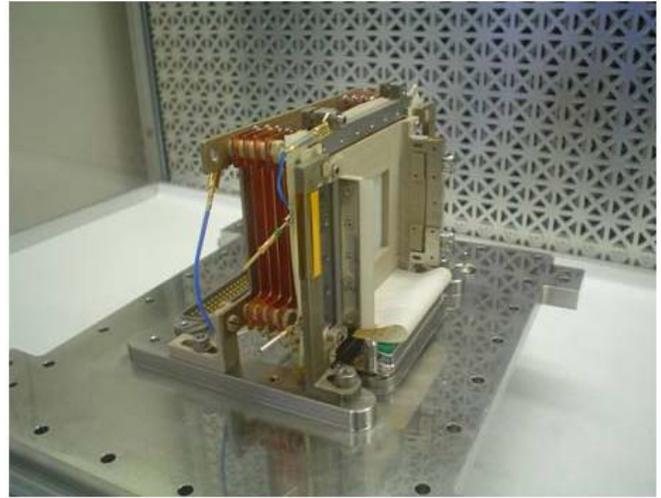

Figure 2. Image of the assembled NITPC showing the strip electrode attachment to the 50 pin D-sub connector [7].

In the strong electric field of the GEM holes the electrons are stripped from the $CH_3NO_2$ ions. The free electrons are then accelerated by the strong electric field inside the hole. Collisions with gas atoms result in a Townsend avalanche that multiplies the single electron into thousands. The electron cloud is then collected on strip electrodes. The response from the 32 charge-sensitive amplifiers is binned into a 32 x 32 pixel$^2$ image centered on the interaction point. Strip number defines one pixel coordinate and arrival time the other. The charge in each time bin is calculated as $q_n \approx V_n - V_{n-1}$, where $V_n$ is the digitized voltage at time bin $n$ [4]. It is important to select ADCs with the appropriate readout speed as this determines the time bin size. The bin size in the time dimension must be matched to the negative ion drift velocity so that square pixels are produced in the photoelectron track image. The pixel width in 1-dimension is determined by strip pitch, 0.121 mm in the NITPC. Measured negative ion drift velocities are reported in [18]. The drift velocity varies with gas composition and electric field. The fixed sample rate in our experiment was 2.5 MHz. The required time bin width varies with the drift velocity. Each time bin is then an accumulation of the charge from a number of samples to produce a bin of the appropriate size. For example, with a drift field of 909 V/cm the drift velocity is 20.1 m/s. The time bin size should then be 6.02 $\mu$s, requiring a binning of 15 samples/bin.

A track image is formed using the strip number where the charge was accumulated and charge arrival time. The position of each pixel in the image is then determined by the strip number and arrival time of the charge packet. Example track images are shown in Figure 3.

#### 2) Implementation

The NITPC utilizes a double GEM to achieve a sufficient detector gain within the operational voltages of the GEM [19]. The GEMs (produced by SciEnergy, [20]) have a 5x5 cm$^2$ active area with 0.070 mm diameter holes on a 0.140 mm hexagonal pitch. The copper electrodes are 0.005 mm thick separated by 0.100 mm liquid crystal polymer (LCP). The GEMs are mechanically tensioned with ~18 N of force before being mounted in titanium frames to match the coefficient of thermal expansion of the LCP. The frames utilize a knife-edge to capture the LCP border and ensure that the GEM does not creep once taut. The GEM frames are separated by an insulator and maintained at different voltages. The GEM separation is 2.6 mm.

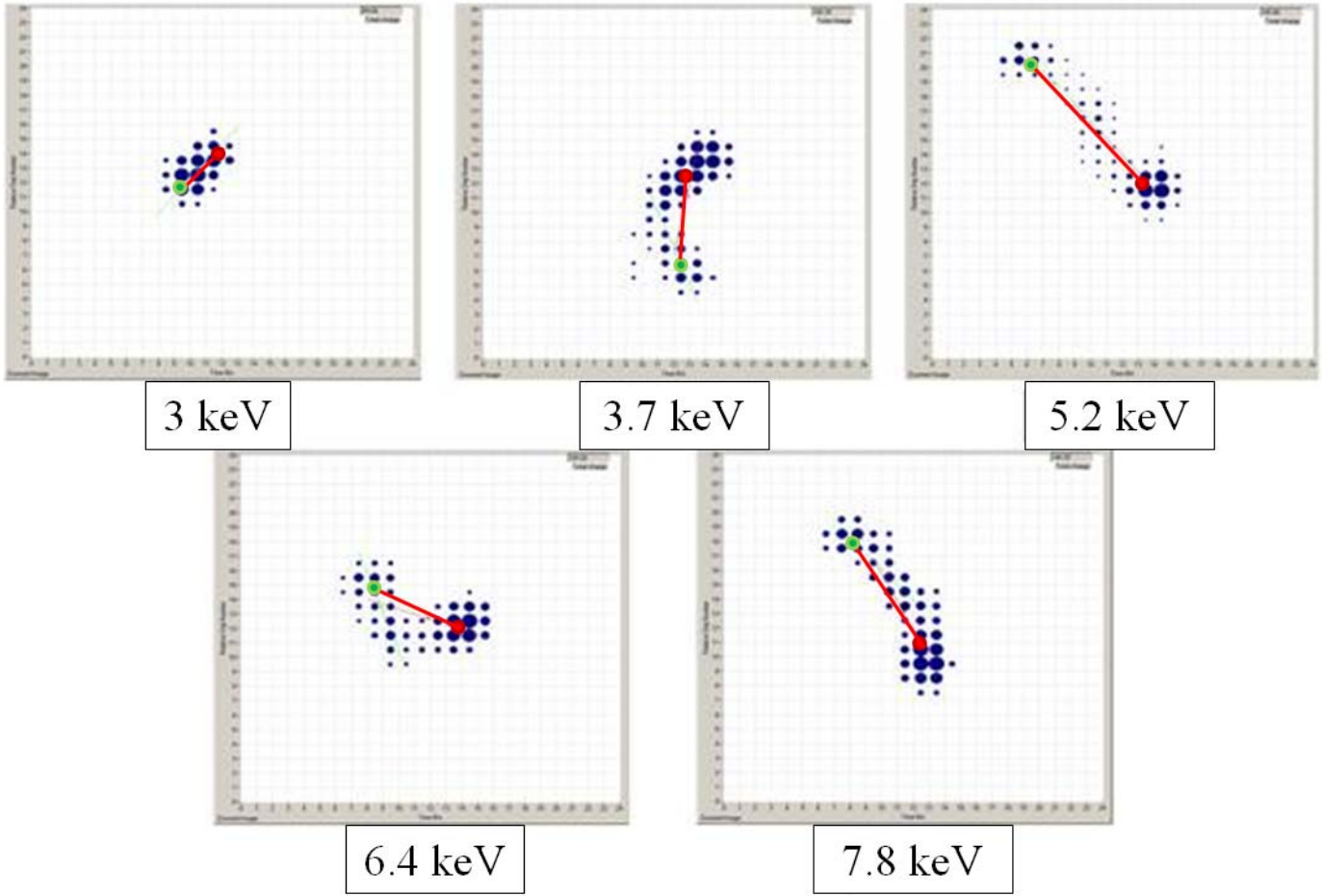

Figure 3. Photoelectron track images in a gas mixture of $Ne+CO_2+CH_3NO_2$. The grid is 0.121 x 0.121 $mm^2$, with the vertical axis corresponding to the strip anode position and the horizontal axis the timing direction. The size of the circle in each pixel represents the relative amount of charge in that pixel. The green circle with the darker center denotes the interaction point, the red solid circle is the barycenter of the densest portion of the track.

The anode copper strip electrodes are on 50 μm LCP and have a pitch of 0.121 mm. The strip orientation is parallel to the vertical axis as shown in Figure 1. Every 32$^{nd}$ strip of the 256 total strips is tied together to minimize the readout instrumentation. The strips are mounted by securing two sides with PEEK frames that have a knife-edge similar to that of the GEM frames. Before the frames are tightened, the strips are pulled taut with a mechanical screw system. The strips wrap around the PEEK frame on each side so that the mounting mechanisms will not interfere with maintaining a narrow gap between the strips and GEM. This gap is 0.5 mm. The output leads of the strips are run down the long portion of the strips. This long "tail" is then connected to a 50D-sub connector. The strips are mounted 0.5 mm beneath the double GEM assembly and the strip-GEM assembly is then attached to a stainless steel flange where the 50D-sub connector interfaces to a 50 pin hermetically sealed feedthrough in the gas enclosure.

The drift electrode is constructed from an iridited (iridited aluminum has a chromate coating that provides corrosion protection. This coating was applied by the engineering branch at NASA GSFC.) Al plate, to prevent the Al and nitromethane from reacting. The field cage consists of five copper plates machined into a C-shape separated by PEEK spacers. Each copper element has a 0.127 mm diameter stainless steel wire stretched across the opening of the C-shape to complete the cage while minimizing the obstruction to incident X-rays. The distance between drift electrode and the top GEM for this experiment was 13 mm.

The top electrode of the GEM closest to the strips is instrumented with an Ortec 142PC charge sensitive amplifier and an Ortec 671 shaping amplifier. The output from the shaping amplifier is used to generate a trigger for the strip electrode readout. Each set of 32 strips is instrumented with a charge sensitive amplifier (Cremat 110). The signals are digitized and read out via USB 2.0 and analyzed using a program written in C code with LabWindows 8 software suite.

Table 1. Operating voltages of the NITPC in $Ne+CO_2+CH_3NO_2$ (260:80:10 Torr) and $CO_2+CH_3NO_2$ (197.5: 15 Torr). *Drift* specifies the voltage applied to the drift electrode. *F, C, and A* refer to the frame, cathode, and anode voltages for the first (1) and second (2) GEM. *Units:* V.

|  | Drift | F1 | C1 | A1 | F2 | C2 | A2 |
|---|---|---|---|---|---|---|---|
| $Ne+CO_2+CH_3NO_2$ | 3115 | 1715 | 1515 | 1030 | 900 | 770 | 320 |
| $CO_2+CH_3NO_2$ | 3150 | 1820 | 1630 | 1100 | 900 | 800 | 300 |

### B. Brookhaven National Laboratory Experimental Setup

The detector performance was tested at the X19A beamline at the Brookhaven National Laboratory (BNL) National Synchrotron Light Source (NSLS) in late November 2010. Beamline X19A provides 2.1-17 keV X-rays with an intensity of approximately $10^{11}$ photons sec$^{-1}$.

Data were collected from the nearly 100 % polarized NSLS beam line, [21], at multiple energies between 3-8 keV. The 1x2 mm$^2$ beam is collimated by two 0.5 mm diameter holes spaced 55 mm apart. This produces a 1 mm diameter beam in the detector and reduces the count rate. Event rates were further limited to approximately 30 Hz by de-tuning the second crystal of the double crystal monochromator. The nominal operating voltages for each gas mixture are reported in Table 1, with variations of the Drift and F1 values to create different drift region electric field strengths. An apparent gas gain of approximately 2000 at the strip anodes was maintained in both gas mixtures.

The data reported here were taken over two days. For each gas mixture, all of the data presented were taken back to back within a few hours of each other and from the same gas fill. Data were also collected at various drift positions in the detector (2, 6, and 10 mm from the amplification region) and with multiple drift electric field strengths (1143, 1407, and 1693 V/cm).

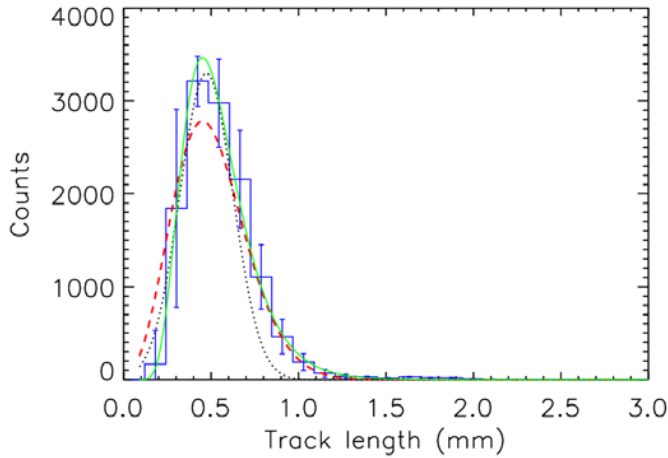

Figure 4. Comparison of photoelectron track length distribution fits to 5.2 keV data in $Ne+CO_2+CH_3NO_2$ (260:80:10 Torr). Three different distributions were fit to the data: lognormal (solid), Poisson (dash), Gaussian (dot). We describe our fitting process and results in detail at the beginning of §IV.

## III. ANALYSIS

### A. Photoelectron track length determination

The track length definition used here measures the distance between barycenters of the two densest regions of charge in the track. This definition was chosen because, for X-ray energies used in this study, it consistently represents the distance from the interaction point to the end of the track where the electron loses most of its charge. The two ends of the track are ill-defined given our resolution and a barycenter of the charge in each is a reasonable approach to estimating where the X-ray interacted and where the photoelectron 'stopped'.

The pixel threshold level is set at 3-times the RMS noise level (~3 e$^-$) and only those pixels with charge greater than this threshold were included in the image. An event cut is performed based on the number of pixels in an image above the threshold. If the image has 15 or fewer pixels above threshold then it is excluded because there are not enough pixels to accurately determine the initial direction of the track.

The interaction point is found by first determining the furthest hit pixel from the track barycenter, where the barycenter coordinates are given by the following:

$$x_b = \frac{\sum_i q_i x_i}{\sum_i q_i} \qquad y_b = \frac{\sum_i q_i y_i}{\sum_i q_i} \qquad (1)$$

The pixel with the most charge within 3 pixels of the furthest hit pixel from the barycenter is called the interaction point. The majority of the charge is deposited at the end of a track, thus the barycenter is always closer to the end of the track and the furthest hit pixel from the barycenter is near the interaction point. The distance from the barycenter to the initial interaction point is then defined as the photoelectron track length [15]. Figure 3 shows sample track images at multiple energies with interaction points, barycenters, and track length estimates.

## IV. RESULTS AND DISCUSSION

The photoelectron track lengths for X-ray energies from 3 – 8 keV were measured in $Ne+CO_2+CH_3NO_2$ and $CO_2+CH_3NO_2$. For each energy, approximately $2x10^4$ qualifying events were recorded and the track length determined as described in §III. Track length histograms were generated at multiple energies for each gas. Two sources of uncertainty were included for the data in each bin ($N$): the standard deviation $\sqrt{N}$ and the effective uncertainty due to the bin width [22], $\Delta N_{eff}$. The smallest bin width is limited by the detector pixel size, 0.121 mm. $\Delta N_{eff}$ is then determined by the following: $\Delta N_{eff,i} = \left|\frac{\partial y_i}{\partial x_i}\right|\delta x$, where $\left|\frac{\partial y_i}{\partial x_i}\right|$ is the slope at any data point, $i$, of an initial fit to the data and $\delta x$ is half the bin width. The total uncertainty is given by each source added in quadrature.

The histograms, with uncertainties as described above, were fitted with Gaussian, Poisson and lognormal distributions. The lognormal distribution gave the best fit to the data in all cases. Figure 4 shows an example track length distribution for 5.2 keV X-rays with all three distributions fit to the data. The lognormal fit has a $\chi^2$ of 4.00 with 10 degrees of freedom (DoF), giving greater than 99.5 % confidence that the data is lognormal. The Gaussian distribution has a $\chi^2$ of 510 with 15 DoF and the Poisson a $\chi^2$ of 272 with 16 DoF, less than 0.05 % confidence of the data being Gaussian or Poisson. The lognormal distribution is defined as follows:

$$lognormal(x) = \frac{A}{x} exp\left(\frac{-\left(\log\left[\frac{x}{x_c}\right]\right)^2}{2w^2}\right) \qquad (2)$$

where the mean track length is given by $x_c$ (mm). Figure 5 shows lognormal distributions with fits for 3 energies in $CO_2+CH_3NO_2$ and $Ne+CO_2+CH_3NO_2$ respectively. Lognormal distributions often arise in nature when a random variable (track length in this case) is the product of numerous independent random variables (photoelectron kinetic energy, distance between elastic & inelastic collisions) combined together.

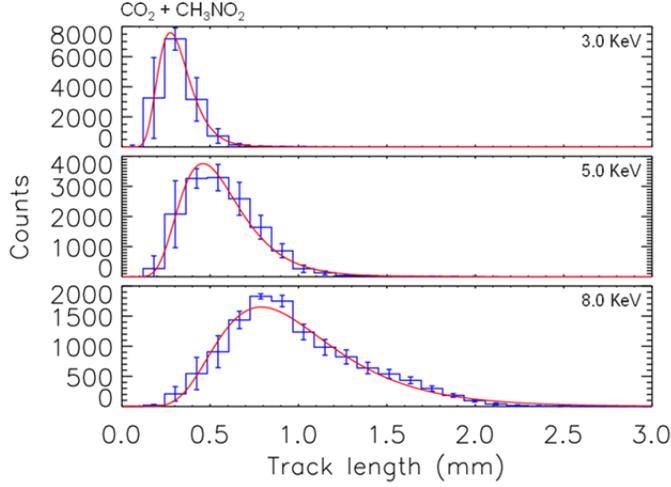

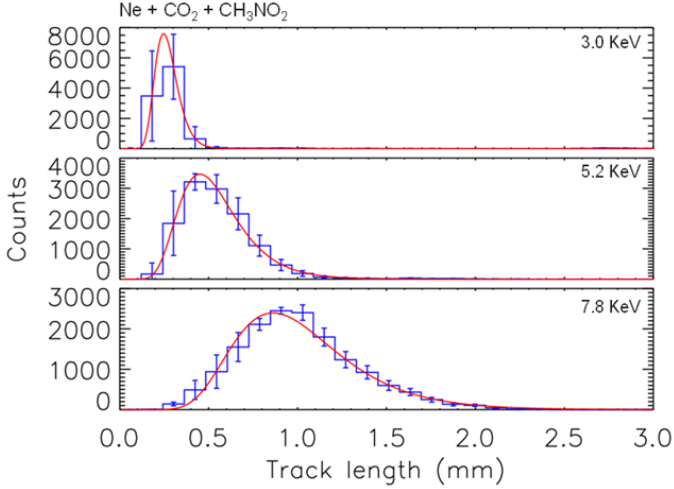

Figure 5. Photoelectron track length data fit with a lognormal distribution in $CO_2+CH_3NO_2$ (197.5:15 Torr) [Top] and $Ne+CO_2+CH_3NO_2$ (260:80:10 Torr) [Bottom]. Showing 1σ errors with an additional element due to bin size. See §IV for a detailed discussion of the uncertainties.

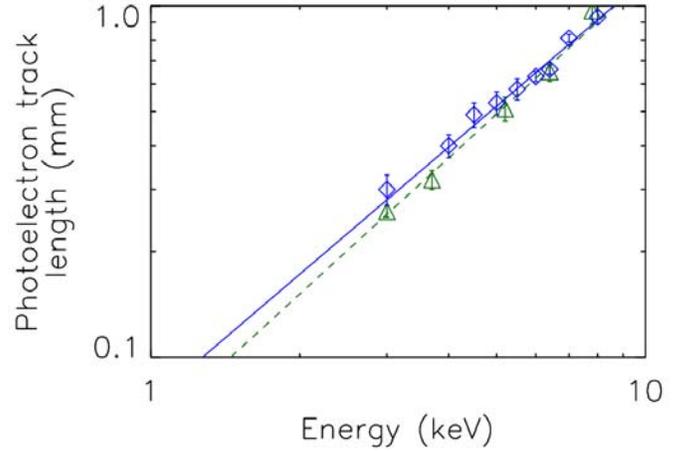

Figure 6. Log-log plot of photoelectron track length as a function of X-ray energy in $CO_2+CH_3NO_2$ (197.5:15 Torr) [blue diamonds, fit is shown as a solid blue line] and $Ne+CO_2+CH_3NO_2$ (260:80:10 Torr) [green triangles, fit is shown as a dashed green line]. The data are best fit with a powerlaw, $f(E) = a(E/E_o)^n$. The fit parameters for each data point shown are given in Table 2. The fit parameters for the power-law fit are reported in Table 3 and results are described in §IV.

Table 2. Lognormal fit parameters for photoelectron track length data points plotted in Figure 6.

| Energy (keV) | $x_{c\ (mm)}$ | $w$ | $\chi^2$ | DoF |
|---|---|---|---|---|
| $CO_2+CH_3NO_2$ | | | | |
| 3 | 0.30±0.03 | 0.32±0.04 | 4.09 | 5 |
| 4 | 0.40±0.03 | 0.40±0.03 | 12.6 | 12 |
| 4.5 | 0.49±0.04 | 0.40±0.02 | 11.6 | 15 |
| 5 | 0.53±0.04 | 0.37±0.02 | 14.5 | 15 |
| 5.5 | 0.58±0.04 | 0.37±0.02 | 13.6 | 15 |
| 6 | 0.63±0.02 | 0.37±0.01 | 16.1 | 19 |
| 6.4 | 0.66±0.02 | 0.40±0.01 | 10.3 | 15 |
| 7 | 0.81±0.02 | 0.38±0.01 | 13.3 | 15 |
| 8 | 0.93±0.04 | 0.41±0.02 | 24.0 | 15 |
| $Ne+CO_2+CH_3NO_2$ | | | | |
| 3 | 0.26±0.01 | 0.25±0.01 | 3.75 | 5 |
| 3.7 | 0.32±0.02 | 0.29±0.03 | 3.20 | 6 |
| 5.2 | 0.51±0.04 | 0.35±0.02 | 4.00 | 10 |
| 6.4 | 0.65±0.04 | 0.34±0.02 | 11.3 | 12 |
| 7.8 | 0.97±0.08 | 0.34±0.01 | 21.4 | 15 |

The mean of the lognormal distribution was used to plot photoelectron track length as a function of X-ray energy. Data are plotted with 1σ uncertainties. For each gas mixture, the track length as a function of energy is well fit by a powerlaw of the form $a(E/E_o)^n$, where $E_o = 1$ keV and $a$ has units of mm. These results are shown in Figure 6. A power-law distribution can be expected for photoelectron track length as a function of energy because the total electron interaction cross-section is itself a power-law distribution at the energies in this study [23], [24]. We find $n$ values of 1.29±0.07 for $Ne+CO_2+CH_3NO_2$ and 1.20±0.09 for $CO_2+CH_3NO_2$. Table 2 provides the lognormal fit parameters for each data point plotted in Figure 6. Table 3 provides the powerlaw fit parameters for each gas mixture.

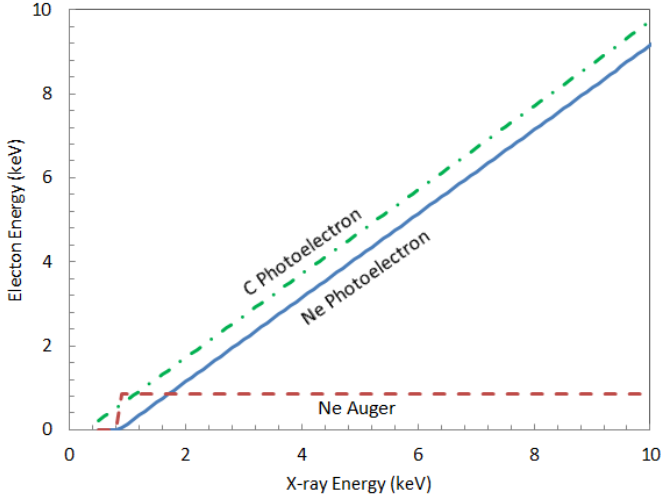

Figure 7. Photoelectron and Auger energy as a function of ionizing X-ray energy for C and Ne. At 3 keV the Ne photoelectron has ~20 % less energy than a C photoelectron due to the presence of a Ne Auger electron. At 8 keV the difference has been reduced to ~7 %.

Table 3. Photoelectron track length as a function of energy. Fit with a powerlaw, $f(x) = ax^n$. Ne+$CO_2$+$CH_3NO_2$ had partial pressures (255:80:15 Torr), $CO_2$+$CH_3NO_2$ (197.5:15 Torr).

|  | $n$ | $a$ (mm) |
| --- | --- | --- |
| Ne+$CO_2$+$CH_3NO_2$ | 1.29±0.07 | 0.062±0.006 |
| $CO_2$+$CH_3NO_2$ | 1.20±0.09 | 0.075±0.012 |

To demonstrate that there was no significant track length dependence on drift velocity we measured track lengths for multiple drift electric field strengths. These measurements were made at an interaction point 6 mm above the detector amplification region. At the highest drift electric field (1693 V/cm) the time for the negative ions to drift into the amplification region would be 48 % longer than at the lowest drift field (1143 V/cm); an increase in track length due to diffusion could occur at a field strength of 1693 V/cm due to this longer drift time. An unpaired t-test comparing the data from each data set gave an average probability of 0.991 for the three comparisons. This suggests that there is no significant difference in track length as a function of drift electric field and the expected difference due to diffusion is below the detection threshold of the experiment. Table 4 shows track length data as a function of electric field and energy in $CO_2$+$CH_3NO_2$.

Track length was also measured as a function of the X-ray interaction point in the detector drift region (referred to as drift position) in order to determine if diffusion would result in significantly different track lengths from different position in the detector. The closest drift position to the amplification region was 2 mm, the furthest 10 mm. At 10 mm, the negative ion drift time is approximately 4 times as long as at 2 mm. The tracks at 10 mm would be expected to be longer than tracks at 2 mm due to diffusion. An unpaired t-test was used to compare the data sets. Table 5 shows the data and t-test results. A comparison of the 2 & 6 cm data have a probability of being from the same distribution of 0.903 and between 2 & 10 mm the probability decreases to 0.787. Examining the data closer we see that the tracks from the 2 mm interaction point are, on average, shorter than the tracks at the 6 mm interaction point by 0.011±0.010 mm and tracks from the 2 mm interaction point are shorter than tracks from the 10 mm interaction point by an average of 0.024±0.008 mm. Considering this result and the t-test results, we conclude that the effects of diffusion are detectable in the NITPC and in Ne+ $CO_2$+$CH_3NO_2$. We find that diffusion adds 3±1e-3 mm to track length per cm of drift distance. Fitting each drift position dependent data set with the $a(E/E_o)^n$ distribution, we find that diffusion does not affect the power in the fit, as each data set gave a result consistent with $n$ = 1.29±0.07 in Ne+ $CO_2$+$CH_3NO_2$.

Table 4. Photoelectron track length (mm) as a function of electric field strength and energy in $CO_2$+$CH_3NO_2$. Each point is from a lognormal fit to a distribution of track lengths. Standard uncertainty is reported.

| | Electric field (V/cm) | | |
| --- | --- | --- | --- |
| Energy (keV) | 1143 | 1407 | 1693 |
| 3 | 0.20±0.01 | 0.21±0.01 | 0.22±0.01 |
| 3.7 | 0.26±0.01 | 0.27±0.01 | 0.27±0.01 |
| 5.2 | 0.45±0.01 | 0.44±0.01 | 0.41±0.01 |
| 6.4 | 0.59±0.01 | 0.61±0.01 | 0.63±0.01 |
| 7.8 | 0.91±0.01 | 0.87±0.02 | 0.87±0.01 |

The two gas mixtures in this experiment have roughly the same density (~520 g/m$^3$) and electron scattering cross-sections (~4e-21 m$^2$ for 1 keV e$^-$) [23], [24]. Photoelectrons of similar energy would be expected to produce similar track length as a function of energy in these gases. We report power-law slopes, $n$, for these gases of 1.29±0.07 for Ne+$CO_2$+$CH_3NO_2$ and 1.20±0.09 for $CO_2$+$CH_3NO_2$. Statistically, these fits are consistent with the same distribution, in agreement with expectations based on electron scattering cross-section and density. However, it is important to note that the lower energy points for Ne+$CO_2$+$CH_3NO_2$ lie more than $1\sigma$ below the power-law fit for $CO_2$+$CH_3NO_2$. We suggest that this difference at low energies is a result of the different atomic make-up of the gases. The mixture with Ne will have approximately 50 % of the interacting X-rays ionizing a Ne atom instead of a C, or O atom due to cross-section and composition. Ionization of C or O will dominate in the other gas mixture.

Figure 7 shows the photoelectron and Auger energy as a function of X-ray energy for C and Ne. For C, the average energy that goes into an Auger electron or fluorescent X-ray is very small compared to the energy going into the photoelectron. In Ne, the Auger electron/fluorescent X-ray path takes a higher average energy from the incident X-ray than for C, resulting in a lower energy photoelectron due to a Ne/X-ray interaction.

At 3 keV the Ne photoelectron has ~20 % less energy than a C photoelectron due to the presence of a Ne Auger electron. At 8 keV the difference has been reduced to ~7 %. This would result in the shorter tracks observed at lower energies in Ne+$CO_2$+$CH_3NO_2$ (~20 % shorter at 3 keV). As the energy of the X-ray increases the Ne photoelectron begins to dominate the Ne Auger and the tracks will become longer. This is also observed in our data and is in agreement with the steeper power-law fit to the Ne+$CO_2$+$CH_3NO_2$ data; the tracks in Ne gas are only ~4 % shorter at 8 keV.

Table 5. Photoelectron track length (mm) as a function of X-ray interaction point (drift position) and energy in Ne+$CO_2$+$CH_3NO_2$. Each point is from a lognormal fit to a distribution of track lengths. Standard uncertainty is reported.

|  | Drift position (cm) | | |
| --- | --- | --- | --- |
| Energy (keV) | 2 | 6 | 10 |
| 3 | 0.24±0.01 | 0.24±0.01 | 0.26±0.01 |
| 4 | 0.33±0.01 | 0.34±0.01 | 0.36±0.01 |
| 4.5 | 0.42±0.01 | 0.42±0.01 | 0.44±0.01 |
| 5 | 0.46±0.01 | 0.47±0.01 | 0.48±0.01 |
| 5.5 | 0.50±0.01 | 0.52±0.01 | 0.53±0.01 |
| 6 | 0.54±0.02 | 0.57±0.02 | 0.58±0.01 |
| 6.4 | 0.59±0.02 | 0.60±0.03 | 0.61±0.01 |
| 7 | 0.73±0.02 | 0.75±0.04 | 0.76±0.01 |
| 8 | 0.85±0.02 | 0.85±0.05 | 0.86±0.02 |

## V. Conclusions

We have completed the first quantitative analysis of photoelectron track length distributions in a gas. Distributions were measured at multiple energies between 3 and 8 keV in two negative ion proportional counter gases, Ne+$CO_2$+$CH_3NO_2$ and $CO_2$+$CH_3NO_2$. The track length distributions are best described as lognormal in all cases. Track length as a function of energy is well fit by a powerlaw of the form $a(E/E_o)^n$ with $n = 1.29\pm0.07$ for Ne+$CO_2$+$CH_3NO_2$ (260:80:10 Torr) and $n = 1.20\pm0.09$ for $CO_2$+$CH_3NO_2$ (197.5:15 Torr). The fit to a powerlaw is also consistent with previous experimental results for photoelectron range in gases and foils.

A NITPC can be used to accurately image low energy photoelectron tracks with a spatial resolution of 0.121 mm in low density gases. Measurements of photoelectron track length in more gas mixtures and at different densities should be performed. As a useful diagnostic for instrument design a wider array of data is now required so that pressure, gas mixture, pixel size and active region dimensions can be optimized. An example of an optimization could be to use the photoelectron track lengths to adjust the resolution of the detector as an improvement in instrument response at low energies. Using our results, we can estimate that in $CO_2$+$CH_3NO_2$ the photoelectron track from a 2 keV X-ray would be approximately 0.170 mm long (Figure 6); this is smaller than the current pixel size of 0.121 x 0.121 $mm^2$ and would not provide accurate initial direction information for the photoelectron. As a result the polarization measurements would be inaccurate. If the NITPC is going to be optimized for low energy X-ray's then pixel size must be reduced or the track length increased (by changing the gas density). Future X-ray polarimeter designs based on a TPC or NITPC will be able to use track length information at the early stages of design to produce instruments which meet sensitivity requirements while optimizing detector parameters.

## Acknowledgment

This work was funded under NASA grant NNX08AF46G at the NASA Goddard Space Flight Center. Beamline X19 A at the Brookhaven National Laboratory National Synchrotron Light Source was utilized to make the measurements presented in this experiment. The authors would like to thank Syed Khalid for his help at BNL. We would also like to thank Israel Moya, Christian Urba, Richard Koenecke and Tracy L Pluchak-Rosnak for their expert technical contributions to the project.